\documentclass[twocolumn,aps,prl,superscriptaddress]{revtex4}

\usepackage{graphicx}
\usepackage{hyperref}

\begin{document}

\title{
Resolving photon number states in a superconducting circuit}

\author{D.~I.~Schuster\footnote{Authors with contributed equally to this work.}}
\affiliation{Departments of Applied Physics and Physics, Yale
University, New Haven, CT 06520}
\author{A.~A.~Houck$^{*}$}
\affiliation{Departments of Applied Physics and Physics, Yale
University, New Haven, CT 06520}
\author{J.~A.~Schreier}
\affiliation{Departments of Applied Physics and Physics, Yale
University, New Haven, CT 06520}
\author{A.~Wallraff}
\affiliation{Departments of Applied Physics and Physics, Yale
University, New Haven, CT 06520}
\affiliation{Department of Physics,
ETH Zurich, CH-8093 Z\"{u}rich, Switzerland}
\author{J.~M.~Gambetta}
\affiliation{Departments of Applied Physics and Physics, Yale
University, New Haven, CT 06520}
\author{A.~Blais}
\affiliation{Departments of Applied Physics and Physics, Yale
University, New Haven, CT 06520}
\affiliation{D\'{e}partement de
Physique et Regroupement Qu\'{e}b\'{e}cois sur les Mat\'{e}riaux de
Pointe, Universit\'{e} de Sherbrooke, Sherbrooke, Qu\'{e}bec,
Canada, J1K 2R1}
\author{L.~Frunzio}
\affiliation{Departments of Applied Physics and Physics, Yale
University, New Haven, CT 06520}
\author{B.~Johnson}
\affiliation{Departments of Applied Physics and Physics, Yale
University, New Haven, CT 06520}
\author{M.~H.~Devoret}
\affiliation{Departments of Applied Physics and Physics, Yale
University, New Haven, CT 06520}
\author{S.~M.~Girvin}
\affiliation{Departments of Applied Physics and Physics, Yale
University, New Haven, CT 06520}
\author{R.~J.~Schoelkopf}
\affiliation{Departments of Applied Physics and Physics, Yale
University, New Haven, CT 06520}
\date{\today}

\begin{abstract}
Electromagnetic signals are always composed of photons, though in
the circuit domain those signals are carried as voltages and
currents on wires, and the discreteness of the photon's energy is
usually not evident. However, by coupling a superconducting qubit to
signals on a microwave transmission line, it is possible to
construct an integrated circuit where the presence or absence of
even a single photon can have a dramatic effect.  This
system\cite{Blais2004} is called circuit quantum electrodynamics
(QED) because it is the circuit equivalent of the atom-photon
interaction in cavity QED. Previously, circuit QED devices were
shown to reach the resonant strong coupling regime, where a single
qubit can absorb and re-emit a single photon many
times\cite{Wallraff2004}. Here, we report a circuit QED experiment
which achieves the strong dispersive limit, a new regime of cavity
QED in which a single photon has a large effect on the qubit or atom
without ever being absorbed.  The hallmark of this strong dispersive
regime is that the qubit transition can be resolved into a separate
spectral line for each photon number state of the microwave field.
The strength of each line is a measure of the probability to find
the corresponding photon number in the cavity.  This effect has been
used to distinguish between coherent and thermal fields and could be
used to create a photon statistics analyzer.  Since no photons are
absorbed by this process, one should be able to generate
non-classical states of light by measurement and perform
qubit-photon conditional logic, the basis of a logic bus for a
quantum computer.

\end{abstract}

\maketitle

\begin{figure}[!bp]

\includegraphics{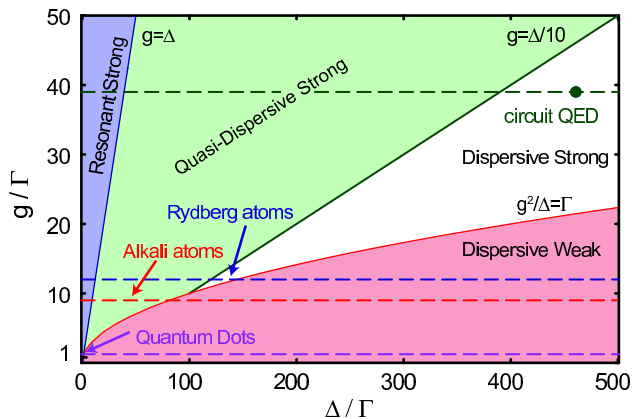}
\caption{A phase diagram for cavity QED. The parameter space is
described by the atom-photon coupling strength, $g$, and the
detuning $\Delta$ between the atom and cavity frequencies,
normalized to the rates of decay represented by
$\Gamma=\max\left[\gamma,\kappa,1/T\right]$. Different cavity QED
systems, including Rydberg atoms, alkali atoms, quantum dots, and
circuit QED, are represented by dashed horizontal lines. The green
circle represents the parameters used in this work. In the blue
region the qubit and cavity are resonant, and undergo vacuum Rabi
oscillations. In the red, weak dispersive, region the ac Stark shift
$g^2/\Delta < \Gamma$ is too small to dispersively resolve
individual photons, but a QND measurement of the qubit can still be
realized by using many photons. In the white region, quantum
non-demolition measurements are in principle possible with
demolition less than $1 \%$, allowing $100$ repeated measurements.
In the green region single photon resolution is possible but
measurements of either the qubit or cavity occupation cause larger
demolition. \label{fig:phasediagram}}
\end{figure}

Cavity QED\cite{Mabuchi2002} is a test-bed system for quantum
optics\cite{Walls2006} that allows investigations into fundamental
questions on quantum measurement and decoherence, and enables
applications such as squeezed light sources and quantum logic gates.
To achieve this, an atom is placed between two mirrors, forming a
cavity that confines the electromagnetic field and enhances the
atom-photon interaction strength.  Cavity QED can be characterized
by this interaction strength, $g$, and the atom-cavity detuning,
$\Delta$, resulting in several regimes which we represent with the
phase diagram in Figure \ref{fig:phasediagram}. Resonance occurs
when the detuning is less than the interaction strength ($\Delta <
g$, blue region in Fig. \ref{fig:phasediagram}), allowing real
excitations to be exchanged between the atom and cavity, resulting
in phenomena such as enhanced spontaneous emission into the cavity
mode (the Purcell effect\cite{Purcell1946}). The \textit{resonant
strong} coupling regime of cavity QED is achieved when the coupling
rate, $g$, is larger than the inverse atom transit time through the
cavity, $1/T$, and the decay rates of the atom, $\gamma$, and
cavity, $\kappa$. In this regime the photon and atom are coherently
coupled, and a single photon is periodically absorbed and re-emitted
(the vacuum Rabi oscillations) at a rate $2g$. Strong coupling has
traditionally been studied in atomic systems using alkali
atoms\cite{THOMPSON1992}, Rydberg atoms\cite{Raimond2001}, or
ions\cite{Leibfried1996,Leibfried1997}. More recently strong
coupling with solid state systems has been achieved with
superconducting
circuits\cite{Wallraff2004,Chiorescu2004,Johansson2006} and
approached in semiconducting quantum
dots\cite{Reithmaier2004,Yoshie2004}. The resonant strong regime of
cavity QED is interesting because the joint system becomes
anharmonic, allowing experiments in non-linear optics and quantum
information at the single photon level.

In the dispersive (off-resonant) limit, the atom-cavity detuning is
larger than the coupling, $\Delta \gg g$ and only virtual photon
exchange is allowed, keeping the atom and photon largely separable
(red and white regions in Fig. \ref{fig:phasediagram}).  The atom
(photon) now acquires only a small photonic (atomic) component of
magnitude $\left(g/\Delta \right)^2$, and an accompanying frequency
shift, $2\chi = 2 g^2/\Delta$.  In this case the system is described
to second order in $g/\Delta$ by the quantum version of the AC Stark
Hamiltonian\cite{Blais2004}:

$$H = \hbar \omega_{\rm{r}} \left(a^{\dag} a+1/2\right)
 + \hbar \omega_{\rm{a}} \, \sigma_{\rm{z}}/2 + \hbar \chi \left(a^{\dag} a +1/2\right)\sigma_{\rm{z}}
 \label{eq:JC}$$

\noindent The first two terms describe a single photon mode as a
harmonic oscillator and an atom or qubit as a two-level pseudo-spin
system. The third term is a dispersive interaction that can be
viewed as either an atom state-dependent shift of the cavity
frequency or a photon number-dependent light shift (the Stark plus
Lamb shifts) of the atom transition frequency. This interaction
means that when the atom state is changed, an energy $2 \hbar \chi$
is added or removed to or from each cavity photon. This interaction
is of particular interest because it commutes with the individual
atom and photon terms, meaning that it is possible to do a quantum
non-demolition\cite{Grangier1998,Caves1980} (QND) measurement of
either the atom state by measuring the phase shift of photons in the
cavity\cite{Wallraff2005} or photon number using the atomic Stark
shift\cite{Brune1994,Schuster2005}.  The demolition of a measurement
is quantified by the probability that in the absence of any other
decay mechanism, a repetition of the measurement will yield a
different result.  To realize a QND measurement, one could drive the
atom at the Stark shifted atom frequency ($\omega_{\rm{a}}+2 n
\chi$), selectively exciting it if there are exactly n photons in
the cavity, and then measure the atom state independently to readout
the result.  In our experiment, the cavity transmission is used to
measure the atom state, so while the interaction is QND, the
detection performed here is not. A more fundamental limitation for
any cavity QED experiment arises from the second order mixture of
the atomic and photonic states, creating a probability,
$(g/\Delta)^2$, that a measurement of photon number will absorb a
photon or a measurement of the atomic state will induce a
transition, demolishing the measured state.

In analogy with the strong resonant case, the \textit{strong
dispersive} limit can be entered when the Stark shift per photon is
much larger than the decoherence rates ($2\chi > \gamma,\kappa,1/T$
white region in Fig. \ref{fig:phasediagram}), while the demolition
remains small $(g/\Delta)^2 \ll 1$. The small number-dependent
frequency shift present in the weak dispersive regime (red region
Fig. \ref{fig:phasediagram}), becomes so large that each photon
number produces a resolvable peak in the atomic transition spectrum,
allowing the measurement in this paper. It has been proposed that
the dispersive photon shift could be used to make a QND measurement
of the photon number state of the cavity using Rydberg
atoms\cite{BRUNE1990}. Previously attainable interaction strengths
required photon number detection experiments to employ absorptive
quantum Rabi oscillations in the resonant regime\cite{Brune1996},
allowing a QND measurement\cite{Nogues1999} restricted to
distinguishing only between zero and one photon.  More recently, a
non-resonant Rydberg atom experiment entered the strong dispersive
limit, measuring the single photon Wigner function with demolition
$(g/\Delta)^2=6 \%$, in principle allowing $\sim 15$ repeated
measurements\cite{Bertet2002}. We present here a circuit QED
experiment clearly demonstrating the strong dispersive regime,
resolving states of up to ten photons, and having demolition
$(g/\Delta)^2 < 1 \%$, which should allow up to $\sim 100$ repeated
QND measurements.

\begin{figure*}
\includegraphics*[width = 1 \textwidth]{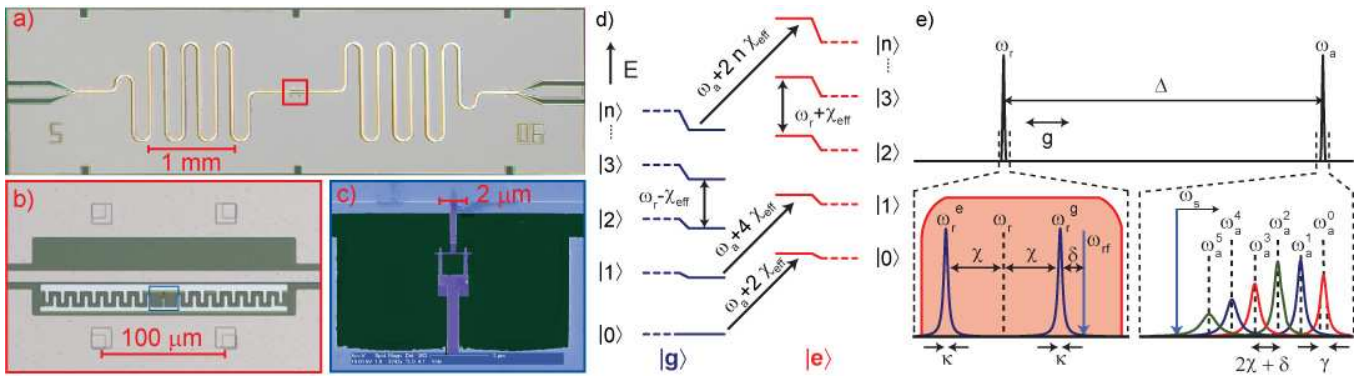}
\caption{Cooper Pair Box (CPB) inside cavity and spectral features
of the circuit QED system. \textbf{a.} An on-chip coplanar waveguide
cavity with resonant frequency $\omega_{\rm{r}}/2\pi=5.7 \,
\rm{GHz}$.
 \textbf{b.} The CPB, placed at a voltage anti-node of the coplanar waveguide
(CPW) cavity (metal is beige, substrate is dark), consists of two
large superconducting islands (light blue) connected by a pair of
Josephson tunnel junctions (purple in \textbf{c}). Both the CPB and
cavity are made from Aluminum, a superconductor at the experiment
temperature, $T=20 \, \rm{mK}$. The transition frequency between the
lowest two CPB levels is $\omega_{\rm{a}}/2\pi \approx \sqrt{8
E_{\rm{J}} E_{\rm{C}}}/h=6.9 \, \rm{GHz}$, where the Josephson
energy $E_{\rm{J}}/h=11.5 \, \rm{GHz}$ and the charging energy
$E_{\rm{C}}/h=e^2/2C_{\Sigma}h=520 \, \rm{MHz}$. Both the large
dipole coupling, $g/2\pi=105 \, \rm{MHz}$, that allows access to the
strong dispersive regime, and the small charging energy are due to
the large geometric capacitance of the box islands to the resonator.
With these parameters the transition frequency from ground to first
excited state is larger by $10 \%$ than the next lowest transition,
allowing the two levels to be addressed uniquely, though higher
levels do contribute dispersive shifts.  Most notably the nearest
level causes a negative effective Stark shift per photon,
$2\chi_{\rm{eff}}=-17 \, \rm{MHz}$, as well as a Lamb shift-like
dressing of the resonator and qubit frequencies.  This dispersive
shift is larger than the linewidths of both the qubit ($\gamma/2
\pi=1.9 \, \rm{MHz}$) and cavity ($\kappa/2\pi=250 \, \rm{kHz}$).
\textbf{d.} Dispersive cavity-qubit energy levels.  Each energy
level in the qubit-cavity Hamiltonian is labeled by the qubit state,
where right is excited $\left|e\right>$, and left is ground,
$\left|g\right>$, while $\left|n\right>$, denotes the number of
photons in the cavity. The dashed portion represents the
qubit/cavity energy levels with no interaction ($g=0$), where the
solid lines show the eigenstates as they are dressed by the
dispersive interaction.  Transitions from $\left|n\right>
\rightarrow \left|n+1\right>$ show the qubit-dependent cavity shift.
Transitions at constant photon number from
$\left|g\right>\left|n\right> \rightarrow
\left|e\right>\left|n\right>$ (left to right) show a photon number
dependent frequency shift $2 n \chi_{\rm{eff}}$. \textbf{e.} Each
transition in \textbf{d} can be measured in the cavity-qubit
spectral response.  The qubit state-dependent spectral response of
the cavity is shown in the bottom left. To measure the qubit state,
and also to populate the cavity in Figs. \ref{fig:numbersplitting}
and \ref{fig:photonstatistics}a-b, the cavity is driven with a
coherent tone at $\omega_{\rm{rf}}$, that is blue detuned from the
cavity by several linewidths to not induce any cavity nonlinearity.
To attain a thermal distribution the cavity was driven with gaussian
noise spanning the cavity according to the red envelope.  The qubit
spectrum is shown at the bottom right, and is detuned from the
cavity by $\Delta/2\pi=1.2 \, \rm{GHz} \gg \,$ $ g/2\pi$.
Measurements are performed by measuring qubit response to being
driven with a spectroscopy tone, $\omega_{\rm{s}}$, by monitoring
transmission at $\omega_{\rm{rf}}$. Because $\chi > \gamma$ each
photon shifts the qubit transition by more than a linewidth giving a
distinct peak for each number of photons in the cavity. The maximum
number of resolvable peaks is determined by $2\chi/\kappa$.
\label{fig:sample}}
\end{figure*}

In circuit QED\cite{Blais2004,Wallraff2005} the ``atom''-photon
interaction is implemented by a Cooper Pair Box
(CPB)\cite{Bouchiat1998}, chosen for its large dipole moment,
capacitively coupled to a full-wave one-dimensional transmission
line resonator. The resonator's reduced transverse dimensions,
microns instead of centimeters (Fig. \ref{fig:sample}), enhance the
energy density a million times over a three-dimensional microwave
cavity.  This large energy density, together with the large
geometric capacitance (dipole moment) of the CPB, yield an
interaction strength that is $g/\omega_{\rm{a,r}} = 2\%$ of the
total photon energy. This coupling, ten-thousand times larger than
currently attainable in atomic systems, allows circuit QED to
overcome the larger decoherence rates present in the solid-state
environment, maintaining $g/\gamma_{\rm{eff}}=40$ possible coherent
vacuum Rabi oscillations in the strong resonant regime, where
$\gamma_{\rm{eff}}=(\gamma+\kappa)/2$.  The equivalent comparison of
the dispersive interaction to decoherence examines the Stark shift
per photon in relation to the qubit decay, $2\chi/\gamma = 6$, and
determines the resolution of photon number peaks. Comparing instead
to the cavity lifetime yields an estimate of the maximum number of
peaks that could possibly be resolved, $2\chi/\kappa = 70$, and
determines the contrast of a qubit measurement by the cavity. These
values of our parameters place the system well into the strong
dispersive regime.

\begin{figure}[!bp]

\includegraphics[scale=1]{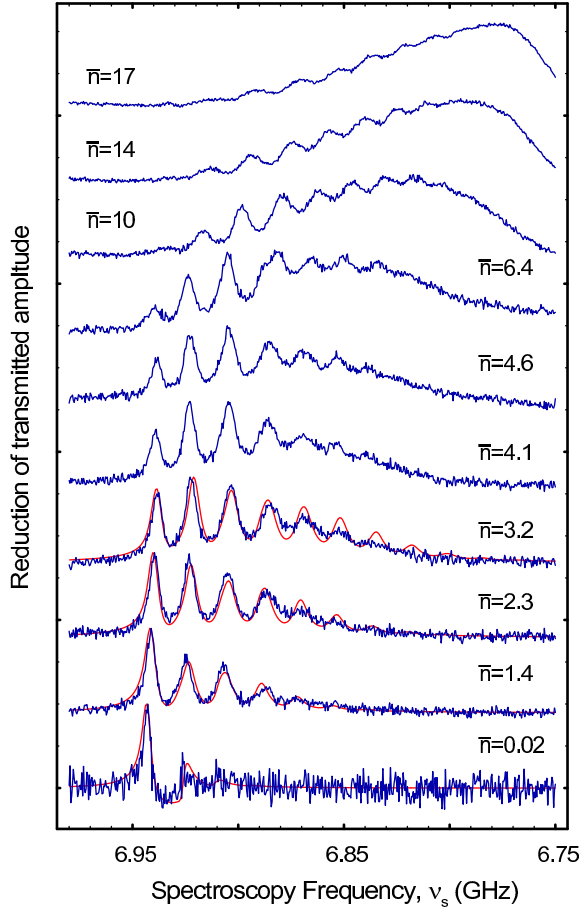}
\caption{Direct spectroscopic observation of quantized cavity photon
number. Qubit spectra with coherent cavity drive at different
average cavity occupations ($\overline{n}$).  The spectra have
resolved peaks corresponding to each photon number.  The peaks are
separated by $2 \chi_{\rm{eff}}/2\pi=-17 \, \rm{MHz}$. Approximately
ten peaks are distinguishable.  The data (blue) is well described by
numerical simulations (red) with all parameters predetermined except
for a single frequency offset, overall power scaling, and background
thermal photon number ($n_{\rm{th}} = 0.1$) used for all traces.
Computational limitations prevented simulations of photon numbers
beyond $\approx 3$. At the lowest power nearly all of the weight is
in the $\left|0\right>$ peak, meaning that the cavity has a
background occupation less than ($n_{\rm{th}}< 0.1$). Peaks broaden
as $\left(n+\overline{n}\right) \kappa$ plus some additional
contributions due to charge noise.  At higher powers the peaks blend
together and the envelope approaches a gaussian shape for a coherent
state. Since $\chi < 0$, spectra are displayed from high to low
frequency, and also have been normalized and offset for clarity.
\label{fig:numbersplitting}}
\end{figure}

\begin{figure}[!bp]
\includegraphics[scale=1.0]{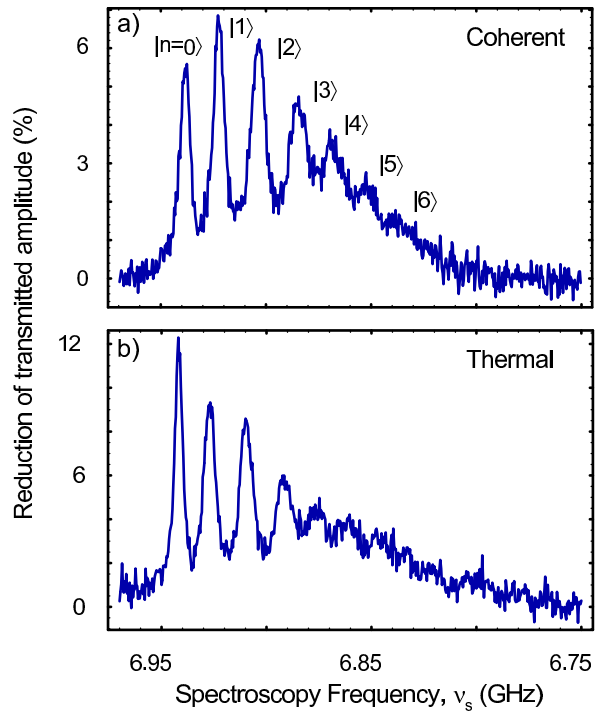}
\caption{Qubit spectrum distinguishes between coherent and thermal
distributions. \textbf{a.} Reduction in transmitted amplitude is
plotted as a proxy for qubit absorption for the case of a coherent
drive with $\overline{n}=3$ photons. \textbf{b.} Spectrum when
cavity is driven with Gaussian white noise approximating a thermal
state also with $\overline{n}=3$.  The coherent spectrum is clearly
non-monotonic and qualitatively consistent with the Poisson
distribution, $P(n) = e^{-\overline{n}} \overline{n}^{n} / n!$,
while the thermal spectrum monotonically decreases consistent with
the Bose-Einstein distribution $P(n) =
\overline{n}^{n}/\left(\overline{n}+1\right)^{n+1}$.
\label{fig:photonstatistics}}
\end{figure}

The photon number dependent frequency shift of the qubit is detected
by performing spectroscopy on the qubit-cavity system (Fig.
\ref{fig:sample}e).  Photons are placed in the cavity by applying a
microwave signal (the cavity tone) at frequency ($\omega_{\rm{rf}}$)
near the cavity resonance (Fig. \ref{fig:sample}e).  A spectrum is
taken by sweeping the frequency ($\omega_{\rm{s}}$) of a second
microwave signal (the spectroscopy tone), which probes the qubit
absorption, without significantly populating the resonator as it is
detuned by many linewidths $\left(\omega_{\rm{s}}-\omega_{\rm{r}}
\gg \kappa \right)$. The detection is completed by exploiting the
dual nature of the qubit-photon coupling, reusing the cavity photons
as a measure of cavity transmission, demonstrated
previously\cite{Blais2004,Schuster2005,Wallraff2005,Wallraff2004} to
measure the qubit excited state population.  The measured
transmission amplitude (Figs.
\ref{fig:numbersplitting}-\ref{fig:photonstatistics}) is an
approximate measure of the actual qubit population, which could in
principle be measured independently.  For clarity the transmission
amplitude in Figures
\ref{fig:numbersplitting}-\ref{fig:photonstatistics} is plotted from
high to low frequency. In order to reduce non-linearities in the
response, the cavity tone was applied at a small detuning from the
resonator frequency when the qubit is in the ground state $\delta /
2\pi=\left(\omega_{\rm{rf}}-\omega^{\rm{g}}_{\rm{r}}\right)/2\pi=2
\, \rm{MHz}$ which also slightly modifies the peak
splitting\cite{Gambetta2006} (Fig. \ref{fig:sample}e).

The measured spectra reveal the quantized nature of the cavity
field, containing a separate peak for each photon number state (Fig.
\ref{fig:numbersplitting})\cite{Gambetta2006,Irish2003}.  These
peaks approximately represent the weight of each Fock state in a
coherent field with mean photon number $\overline{n}$, which is
varied from zero to seventeen photons. At the lowest photon powers,
nearly all of the weight is in the first peak, corresponding to no
photons in the cavity, and confirming that the background cavity
occupancy is $n_{\rm{th}} < 0.1$. As the input power is increased,
more photon number peaks can be resolved and the mean of the
distribution shifts proportional to $\overline{n}$. The data agree
well with numerical solutions at low powers (solid lines in Fig.
\ref{fig:numbersplitting}) to the Markovian Master
equation\cite{Walls2006,Gambetta2006} with three damping sources,
namely the loss of photons at rate $\kappa /2\pi=250 \, \rm{kHz}$,
energy relaxation in the qubit at rate $\gamma_1/2 \pi= 1.8 \,
\rm{MHz}$ and the qubit dephasing rate $\gamma_{\phi}/2 \pi= 1.0 \,
\rm{MHz}$. However adequate numerical modeling of this strongly
coupled system at higher photon numbers is quite difficult and has
not yet been achieved.

In earlier work\cite{Brune1994,Schuster2005} in the weak dispersive
limit ($\chi/\gamma<1$), the measured linewidth resulted from an
ensemble of Stark shifts blurring the transition, while here in the
strong limit ($\chi/\gamma>1)$ each member of the ensemble is
individually resolved. In the spectra measured here (Fig.
\ref{fig:numbersplitting}), the linewidth of a single peak can be
much less than the frequency spread of the ensemble, but changes in
photon number during a single measurement can still completely
dephase the qubit.  Taking this into account yields a predicted
photon number dependent linewidth, $\gamma_{\rm{n}}= \gamma/2 +
\gamma_{\phi}+\left(\overline{n}+n \right) \kappa/2$ for the
$n^{th}$ peak\cite{Gambetta2006}. The lowest power peak (in the
$\overline{n}=0.02$ trace) corresponds to zero photons and measures
the unbroadened linewidth, $\gamma_{0}/2\pi = 1.9 \, \rm{MHz}$. When
$\overline{n} = 2\chi_{\rm{eff}} / \kappa$ the peaks should begin to
overlap once more, returning the system to the classical field
regime. If this effect were the only limitation, we might hope to
count as many as $70$ photon number peaks before they merge. In
practice the higher number peaks are also more sensitive to charge
fluctuations in the Cooper pair box, which limits us to about $10$
resolvable photon states in this measurement.

The relative area under each peak in the transmission amplitude
(Fig. \ref{fig:photonstatistics}) contains information about the
photon statistics of the cavity field. We can compare two cases
having the same average cavity occupation ($\overline{n}\sim3$), but
containing either a coherent field (Fig.
\ref{fig:photonstatistics}a) or a thermal field (Fig.
\ref{fig:photonstatistics}b).   To create the thermal field,
gaussian noise was added in a wide band around the cavity (red in
Fig. \ref{fig:sample}e).  The coherent and thermal states are
clearly distinguishable, with the weights of the peaks being
non-monotonic for a coherent distribution while the thermal
distributions were monotonically decreasing for all noise
intensities measured.  However for the sample parameters and
measurement protocols used here, several effects prevent
quantitative extraction of photon number probabilities from the
data.  First the inhomogeneous broadening of the higher number peaks
due to charge noise prevents independent extraction of their areas.
Additionally, though it has been analytically shown that in the
qubit absorption spectrum should accurately represent the cavity
photon statistics\cite{Gambetta2006}, this experiment did not have
an independent means to measure the qubit, and there are
imperfections in mapping the qubit spectrum onto the cavity
transmission. Finally, numerical simulations show that spectroscopic
driving of the qubit results in complex dynamics which squeezes the
cavity photon number, pointing to a path to create exotic states of
light, but also obscuring the initial photon statistics. The
measured data is consistent with numerical predictions which do take
into account such squeezing effects (see Fig.
\ref{fig:numbersplitting}) for photon numbers ($\overline{n} \leq
3$) which we could simulate.  While these effects are large in the
present experiment, an independent measurement of the qubit could be
introduced using a second cavity or Josephson-bifurcation
amplifier\cite{Siddiqi2006}, allowing the realization of a
quantitative photon statistics analyzer.  Previous experiments have
also measured analogous statistics of other Bosonic systems
including phonons in an ion trap\cite{Leibfried1996,Leibfried1997},
excitations in a single electron cyclotron
oscillator\cite{Peil1999}, and the number of atoms in a
Bose-Einstein condensate passing through a cavity\cite{Ottl2005}.

The results obtained here also suggest a method for photon-qubit
conditional logic. The qubit response is now strongly dependent on
the number of photons in the cavity. For example, a controlled-not
(CNOT) gate between a photon and qubit could be implemented by
applying a $\pi$ control pulse at the frequency corresponding to one
photon in the cavity. This would flip the qubit if there were
exactly one photon in the cavity, but do nothing for all other
number states. Since the qubit does not absorb the cavity photon,
the number is unchanged after the operation and could be used to
entangle with distant qubits.  A photon number based gate is
analogous to the phonon common mode coupling used in
ion-traps\cite{Monroe1995}, but since the photons travel along
transmission lines and not through qubits themselves, many qubits
can be placed in a single wavelength, and the photons could be sent
to distant qubits, including those in other cavities.

The observation of resolved photon number peaks in the qubit
spectrum demonstrates a new regime for cavity QED systems, the
strong dispersive limit. Measurement of the spectrum directly
reveals the discrete nature of the microwave field inside the
on-chip cavity.  The qubit spectrum is used to distinguish field
states with different photon statistics.  Further exploitation of
this exceptionally large vacuum Rabi coupling should enable quantum
computing using transmission line cavities as a quantum bus, and
allow preparation of quantum states of light for use in quantum
communication and non-linear optics.


\begin{thebibliography}{10}

\bibitem{Bertet2002}
P.~Bertet, A.~Auffeves, P.~Maioli, S.~Osnaghi, T.~Meunier, M.~Brune,
J.~M.
  Raimond, and S.~Haroche.
\newblock Direct measurement of the {W}igner function of a one-photon {F}ock
  state in a cavity.
\newblock {\em Physical Review Letters}, 89(20):200402, November 2002.

\bibitem{Blais2004}
A.~Blais, R.S. Huang, A.~Wallraff, S.~Girvin, and R.~J. Schoelkopf.
\newblock Cavity quantum electrodynamics for superconducting electrical
  circuits: an architecture for quantum computation.
\newblock {\em Physical Review A}, 69:062320, 2004.

\bibitem{Bouchiat1998}
V.~Bouchiat, D.~Vion, P.~Joyez, D.~Esteve, and M.~H. Devoret.
\newblock Quantum coherence with a single {C}ooper pair.
\newblock {\em Physica Scripta}, T76:165--170, 1998.

\bibitem{BRUNE1990}
M.~Brune, S.~Haroche, V.~Lefevre, J.~M. Raimond, and N.~Zagury.
\newblock Quantum nondemolition measurement of small photon numbers by
  {R}ydberg-atom phase-sensitive detection.
\newblock {\em Physical Review Letters}, 65(8):976--979, August 1990.

\bibitem{Brune1994}
M.~Brune, P.~Nussenzveig, F.~Schmidtkaler, F.~Bernardot, A.~Maali,
J.~M.
  Raimond, and S.~Haroche.
\newblock From {L}amb shift to light shifts: vacuum and subphoton cavity fields
  measured by atomic phase-sensitive detection.
\newblock {\em Physical Review Letters}, 72(21):3339--3342, May 1994.

\bibitem{Brune1996}
M.~Brune, F.~SchmidtKaler, A.~Maali, J.~Dreyer, E.~Hagley, J.~M.
Raimond, and
  S.~Haroche.
\newblock Quantum {R}abi oscillation: A direct test of field quantization in a
  cavity.
\newblock {\em Physical Review Letters}, 76(11):1800--1803, March 1996.

\bibitem{Caves1980}
C.~M. Caves, K.~S. Thorne, R.~W.~P. Drever, V.~D. Sandberg, and
M.~Zimmermann.
\newblock On the measurement of a weak classical force coupled to a
  quantum-mechanical oscillator.
\newblock {\em Reviews of Modern Physics}, 52(2):341--392, 1980.

\bibitem{Chiorescu2004}
I.~Chiorescu, P.~Bertet, K.~Semba, Y.~Nakamura, C.~J. Harmans, and
J.~E. Mooij.
\newblock Coherent dynamics of a flux qubit coupled to a harmonic oscillator.
\newblock {\em Nature}, 431(7005):159--162, September 2004.

\bibitem{Gambetta2006}
J.~Gambetta, Blais A., Schuster~D. I., A.~Wallraff, L.~Frunzio,
J.~Majer, S.~M.
  Girvin, and R.~J. Schoelkopf.
\newblock Qubit-photon interactions in a cavity: Measurement induced dephasing
  and number splitting.
\newblock {\em Submitted to Physical Review A}, 2006.

\bibitem{Grangier1998}
P.~Grangier, J.~A. Levenson, and J.~P. Poizat.
\newblock Quantum non-demolition measurements in optics.
\newblock {\em Nature}, 396(6711):537--542, December 1998.

\bibitem{Irish2003}
E.~K. Irish and K.~Schwab.
\newblock Quantum measurement of a coupled nanomechanical resonator - {C}ooper
  pair box system.
\newblock {\em Physical Review B}, 68(15):155311, October 2003.

\bibitem{Johansson2006}
J.~Johansson, S.~Saito, T.~Meno, H.~Nakano, M.~Ueda, K.~Semba, and
  H.~Takayanagi.
\newblock Vacuum {R}abi oscillations in a macroscopic superconducting qubit
  {LC} oscillator system.
\newblock {\em Physical Review Letters}, 96(12):127006, March 2006.

\bibitem{Leibfried1996}
D.~Leibfried, D.~M. Meekhof, B.~E. King, C.~Monroe, W.~M. Itano, and
D.~J.
  Wineland.
\newblock Experimental determination of the motional quantum state of a trapped
  atom.
\newblock {\em Physical Review Letters}, 77(21):4281--4285, November 1996.

\bibitem{Leibfried1997}
D.~Leibfried, D.~M. Meekhof, C.~Monroe, B.~E. King, W.~M. Itano, and
D.~J.
  Wineland.
\newblock Experimental preparation and measurement of quantum states of motion
  of a trapped atom.
\newblock {\em Journal of Modern Optics}, 44(11-12):2485--2505, 1997.

\bibitem{Mabuchi2002}
H.~Mabuchi and A.~C. Doherty.
\newblock Cavity quantum electrodynamics: Coherence in context.
\newblock {\em Science}, 298(5597):1372--1377, November 2002.

\bibitem{Monroe1995}
C.~Monroe, D.~M. Meekhof, B.~E. King, W.~M. Itano, and D.~J.
Wineland.
\newblock Demonstration of a fundamental quantum logic gate.
\newblock {\em Physical Review Letters}, 75(25):4714--4717, December 1995.

\bibitem{Nogues1999}
G.~Nogues, A.~Rauschenbeutel, S.~Osnaghi, M.~Brune, J.~M. Raimond,
and
  S.~Haroche.
\newblock Seeing a single photon without destroying it.
\newblock {\em Nature}, 400(6741):239--242, July 1999.

\bibitem{Ottl2005}
A.~Ottl, S.~Ritter, M.~Kohl, and T.~Esslinger.
\newblock Correlations and counting statistics of an atom laser.
\newblock {\em Physical Review Letters}, 95(9):090404, August 2005.

\bibitem{Peil1999}
S.~Peil and G.~Gabrielse.
\newblock Observing the quantum limit of an electron cyclotron: {QND}
  measurements of quantum jumps between {F}ock states.
\newblock {\em Physical Review Letters}, 83(7):1287--1290, August 1999.

\bibitem{Purcell1946}
E.~M. Purcell.
\newblock Spontaneous emission probabilities at radio frequencies.
\newblock {\em Physical Review}, 69(11-1):681--681, 1946.

\bibitem{Raimond2001}
J.~M. Raimond, M.~Brune, and S.~Haroche.
\newblock Manipulating quantum entanglement with atoms and photons in a cavity.
\newblock {\em Reviews of Modern Physics}, 73(3):565--582, July 2001.

\bibitem{Reithmaier2004}
J.~P. Reithmaier, G.~Sek, A.~Loffler, C.~Hofmann, S.~Kuhn,
S.~Reitzenstein,
  L.~V. Keldysh, V.~D. Kulakovskii, T.~L. Reinecke, and A.~Forchel.
\newblock Strong coupling in a single quantum dot-semiconductor microcavity
  system.
\newblock {\em Nature}, 432(7014):197--200, November 2004.

\bibitem{Schuster2005}
D.~I. Schuster, A.~Wallraff, A.~Blais, R.~S. Huang, Majer J.,
S.~Girvin, and
  R.~Schoelkopf.
\newblock {AC}-{S}tark shift and dephasing of a superconducting qubit strongly
  coupled to a cavity field.
\newblock {\em Physical Review Letters}, 94:123602, 2005.

\bibitem{Siddiqi2006}
I.~Siddiqi, R.~Vijay, M.~Metcalfe, E.~Boaknin, L.~Frunzio, R.~J.
Schoelkopf,
  and M.~H. Devoret.
\newblock Dispersive measurements of superconducting qubit coherence with a
  fast latching readout.
\newblock {\em Physical Review B}, 73(5):054510, February 2006.

\bibitem{THOMPSON1992}
R.~J. Thompson, G.~Rempe, and H.~J. Kimble.
\newblock Observation of normal-mode splitting for an atom in an optical
  cavity.
\newblock {\em Physical Review Letters}, 68(8):1132--1135, February 1992.

\bibitem{Wallraff2005}
A.~Wallraff, D.~I. Schuster, A.~Blais, L.~Frunzio, J.~Majer, S.~M.
Girvin, and
  R.~J. Schoelkopf.
\newblock Approaching unit visibility for control of a superconducting qubit
  with dispersive readout.
\newblock {\em Physical Review Letters}, 95:060501, 2005.

\bibitem{Wallraff2004}
A.~Wallraff, D.~I. Schuster, A.~Blais, R.-S. Huang, J.~Majer,
S.~Kumar, S.~M.
  Girvin, and R.~Schoelkopf.
\newblock Circuit quantum electrodynamics: Coherent coupling of a single photon
  to a {C}ooper pair box.
\newblock {\em Nature}, 431:162, 2004.

\bibitem{Walls2006}
D.~F. Walls and G.~J. Milburn.
\newblock {\em Quantum optics}.
\newblock Springer, 2006.

\bibitem{Yoshie2004}
T.~Yoshie, A.~Scherer, J.~Hendrickson, G.~Khitrova, H.~M. Gibbs,
G.~Rupper,
  C.~Ell, O.~B. Shchekin, and D.~G. Deppe.
\newblock Vacuum {R}abi splitting with a single quantum dot in a photonic
  crystal nanocavity.
\newblock {\em Nature}, 432(7014):200--203, November 2004.

\end{thebibliography}



\begin{acknowledgments}
This work was supported in part by the National Security Agency
   under the Army Research Office, the NSF,
   the W. M. Keck Foundation, and Yale University.
   A.H. would like to acknowledge support from Yale
   University via a Quantum Information and Mesoscopic Physics
   Fellowship.  A.B. was supported by NSERC, CIAR and FQRNT.  Numerical simulations were performed on a RQCHP cluster.
 The authors declare that they have no competing financial interests.
 Correspondence and requests for materials should be addressed to Rob Schoelkopf~(email:Robert.Schoelkopf@yale.edu).
\end{acknowledgments}


\end{document}